\input epsf

\magnification\magstephalf
\overfullrule 0pt
\def\gsim{\raise.3ex\hbox{$\;>$\kern-.75em\lower1ex\hbox{$\sim$}$\;$}}

\font\rfont=cmr10 at 10 true pt
\def\ref#1{$^{\hbox{\rfont {[#1]}}}$}


\font\fourteenbf=cmbx12 scaled\magstep1

\font\tenbfit=cmbxti10
\font\sevenbfit=cmbxti10 at 7pt
\font\fivebfit=cmbxti10 at 5pt
\newfam\bfitfam 
\textfont\bfitfam=\tenbfit  \scriptfont\bfitfam=\sevenbfit
\scriptscriptfont\bfitfam=\fivebfit

\font\eightit=cmti8

\font\tenbfit=cmbxti10
\font\sevenbfit=cmbxti10 at 7pt
\font\fivebfit=cmbxti10 at 5pt
\newfam\bfitfam 
\textfont\bfitfam=\tenbfit  \scriptfont\bfitfam=\sevenbfit
\scriptscriptfont\bfitfam=\fivebfit

\font\tenbit=cmmib10
\newfam\bitfam
\textfont\bitfam=\tenbit%

\font\tenmbf=cmbx10
\font\sevenmbf=cmbx7
\font\fivembf=cmbx5
\newfam\mbffam
\textfont\mbffam=\tenmbf \scriptfont\mbffam=\sevenmbf
\scriptscriptfont\mbffam=\fivembf

\font\tenbsy=cmbsy10
\newfam\bsyfam 
\textfont\bsyfam=\tenbsy%


\def\pmb#1{\setbox0=\hbox{#1}
 \kern.05em\copy0\kern-\wd0 \kern-.025em\raise.0433em\box0 }

\def\slash{/\kern-.5em}

\def \half {{\textstyle {1 \over 2}}}

 %


\def\boxit#1{\vbox{\hrule\hbox{\vrule\kern1pt\vbox
{\kern1pt#1\kern1pt}\kern1pt\vrule}\hrule}}

\def\h{\hfill\break}
\parskip=6pt
\parindent=0pt
\hsize=17truecm\hoffset=-5truemm
\vsize=23truecm
\def\footnoterule{\kern-3pt
\hrule width 17truecm \kern 2.6pt}


\catcode`\@=11 

\def\nolabels{\def\wrlabeL##1{}\def\eqlabeL##1{}\def\reflabeL##1{}}
\def\writelabels{\def\wrlabeL##1{\leavevmode\vadjust{\rlap{\smash%
{\line{{\escapechar=` \hfill\rlap{\sevenrm\hskip.03in\string##1}}}}}}}%
\def\eqlabeL##1{{\escapechar-1\rlap{\sevenrm\hskip.05in\string##1}}}%
\def\reflabeL##1{\noexpand\llap{\noexpand\sevenrm\string\string\string##1}}}
\nolabels
\global\newcount\refno \global\refno=1
\newwrite\rfile
\def\defref{$^{{\hbox{\rfont [\the\refno]}}}$\nref}
\def\nref#1{\xdef#1{\the\refno}\writedef{#1\leftbracket#1}%
\ifnum\refno=1\immediate\openout\rfile=refs.tmp\fi
\global\advance\refno by1\chardef\wfile=\rfile\immediate
\write\rfile{\noexpand\item{#1\ }\reflabeL{#1\hskip.31in}\pctsign}\findarg}
\def\findarg#1#{\begingroup\obeylines\newlinechar=`\^^M\pass@rg}
{\obeylines\gdef\pass@rg#1{\writ@line\relax #1^^M\hbox{}^^M}%
\gdef\writ@line#1^^M{\expandafter\toks0\expandafter{\striprel@x #1}%
\edef\next{\the\toks0}\ifx\next\em@rk\let\next=\endgroup\else\ifx\next\empty%
\else\immediate\write\wfile{\the\toks0}\fi\let\next=\writ@line\fi\next\relax}}
\def\striprel@x#1{} \def\em@rk{\hbox{}}
\def\lref{\begingroup\obeylines\lr@f}
\def\lr@f#1#2{\gdef#1{\defref#1{#2}}\endgroup\unskip}
\newwrite\lfile
{\escapechar-1\xdef\pctsign{\string\%}\xdef\leftbracket{\string\{}
\xdef\rightbracket{\string\}}}

\def\writestop{\def\writestoppt{\immediate\write\lfile{\string\p
ageno%
\the\pageno\string\startrefs\leftbracket\the\refno\rightbracket%
\string\def\string\secsym\leftbracket\secsym\rightbracket%
\string\secno\the\secno\string\meqno\the\meqno}\immediate\closeout\lfile}}
\def\writestoppt{}\def\writedef#1{}
\catcode`\@=12 
\centerline{\fourteenbf The proton's gluon distribution}\vskip 1truemm
\vskip 8pt
\centerline{A Donnachie}
\centerline{Department of Physics, Manchester University}
\centerline{sandy.donnachie@man.ac.uk}
\vskip 5pt
\centerline{P V Landshoff}
\centerline{DAMTP, Cambridge University}
\centerline{pvl@damtp.cam.ac.uk}
\bigskip
{\bf Abstract}

The gluon distribution is dominated by the hard pomeron at small $x$
and all $Q^2$, with no soft-pomeron contribution. This describes well
not only the DGLAP evolution of
the hard-pomeron part of $F_2(x,Q^2)$, but also charm photoproduction and
electroproduction, and is consistent with what is known about
the longitudinal structure function.

\vskip 15truemm

There is still no agreed fundamental explanation for HERA's striking 
discovery, the surprisingly large rise with increasing $1/x$ of the 
proton structure function $F_2(x,Q^2)$. This rise 
is seen to become more marked as $Q^2$ increases. The conventional 
view\defref\mrs{
A D Martin, R G Roberts, W J Stirling and R S Thorne,
European Physical Journal C23 (2002) 73
}\defref\cteq{
CTEQ Collaboration: J Pumplin et al, JHEP 0207 (2002) 012
}
is that the steeply-rising component is absent at small $Q^2$
and as $Q^2$ increases it is generated through pQCD evolution.
We have argued\defref\cudell{
J R Cudell, A Donnachie and P V Landshoff,
Physics Letters B448 (1999) 281
}
that this view is mathematically suspect, since it relies on an 
expansion of the splitting matrix which is likely to be unsafe\defref\hautmann{
S Catani and F Hautmann, Nuclear Physics B427 (1994) 475
}
because it induces singularities that are almost certainly not present in
the exact matrix.
Instead, therefore, we maintain that a rapidly-rising term is present
already at small $Q^2$, so that pQCD evolution does not generate this
term, but merely makes it become more prominent as $Q^2$ increases.

Our view is supported by ZEUS measurements\defref\zeusc{
ZEUS collaboration: J Breitweg et al, European Physical Journal C12 (2000) 35
}
of the charm structure function $F_2^c(x,Q^2)$, which already displays
a strong rise with increasing $1/x$ even at small $Q^2$. Figure 1
shows the data at $Q^2=1.8$ GeV$^2$; the lines are the conventional
fit\ref{\mrs} to the data (MRST) and our own calculation.  Indeed,
HERA data for photoproduction and electroproduction of charm have the striking
property\defref\dlcharm{
A Donnachie and P V Landshoff, Physics Letters B470 (1999) 243
}
that  at each fixed $Q^2$ they vary 
as the same power $x^{-\epsilon_0}$, with $\epsilon_0\approx 0.4$.
This behaviour is not widely appreciated as the data are normally
shown on a log-linear plot rather than a log-log plot.  By definition, we say
that it is associated with the exchange of an object known as the hard 
pomeron\defref\book{
A Donnachie, H G Dosch, P V Landshoff and O Nachtmann, {\sl Pomeron
Physics and QCD}, Cambridge University Press (2002)
}.
A term with
the same power is present also in the light-quark contribution
to the proton structure function. To a good approximation, the hard-pomeron
coupling
to the charmed quark is found to have the same strength as to each of the
light quarks. Once this assumption of flavour blindness
is made\defref\twopom{
A Donnachie and P V Landshoff, Physics Letters B518 (2001) 63
}
the hard-pomeron component of the complete structure function
$F_2(x,Q^2)$ immediately provides a successful zero-parameter description
of its charm component $F_2^c(x,Q^2)$ at small $x$:
$$
F_2^c(x,Q^2)=A_c~(Q^2)^{1+\epsilon_0}~(1+Q^2/Q_0^2)^
{-1-\epsilon_0/2}~x^{-\epsilon_0}
\eqno(1)
$$
with $Q_0\approx 3$ GeV and $A_c\approx 6\times 10^{-4}$.
We define the charm cross section
$$
\sigma^c(W)={4\pi^2\alpha_{\hbox{{\fiverm EM}}}\over Q^2}F_2^c(x,Q^2)
\Big|_{x=Q^2/(W^2+Q^2)}
$$
With (1),
$$
\sigma^c(W)=0.066\, W^{2\epsilon_0}/(1+Q^2/9.1)^{1+\epsilon_0/2}
\eqno(2)
$$
where the units are $\mu$b and GeV. This expression corresponds
to the thin lines in the plots of figure 2. These lines are not a fit
to these data;  they are calculated from the fit to the hard-pomeron
component of  the complete structure function $F_2(x,Q^2)$,
making the flavour-blindness assumtion.
\topinsert
\centerline{\epsfxsize=0.5\hsize\epsfbox[95 610 330 755]{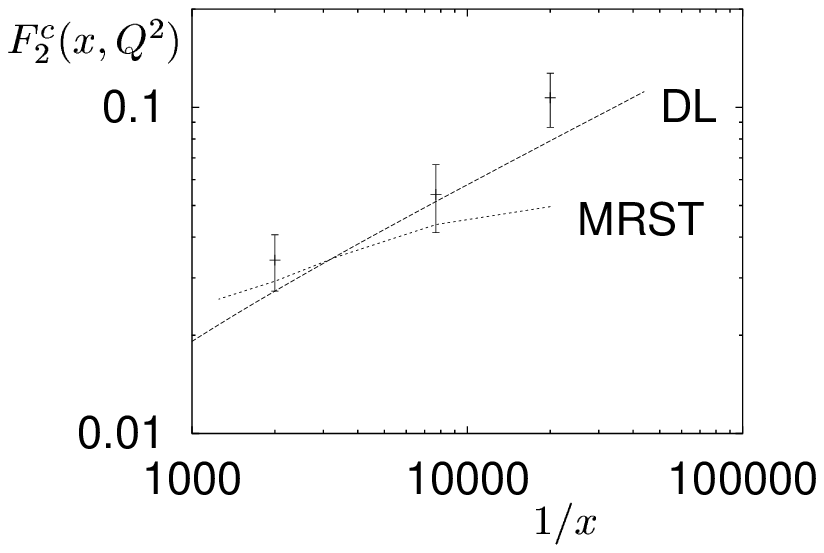}}
\vskip 1truemm
Figure 1: charm structure function data\ref{\zeusc}
at $Q^2=1.8$ GeV$^2$
with curves from MRST\ref{\mrs} and\ref{\twopom}
the form (1)
\endinsert

\topinsert
\epsfxsize=\hsize\epsfbox[70 370 540 770]{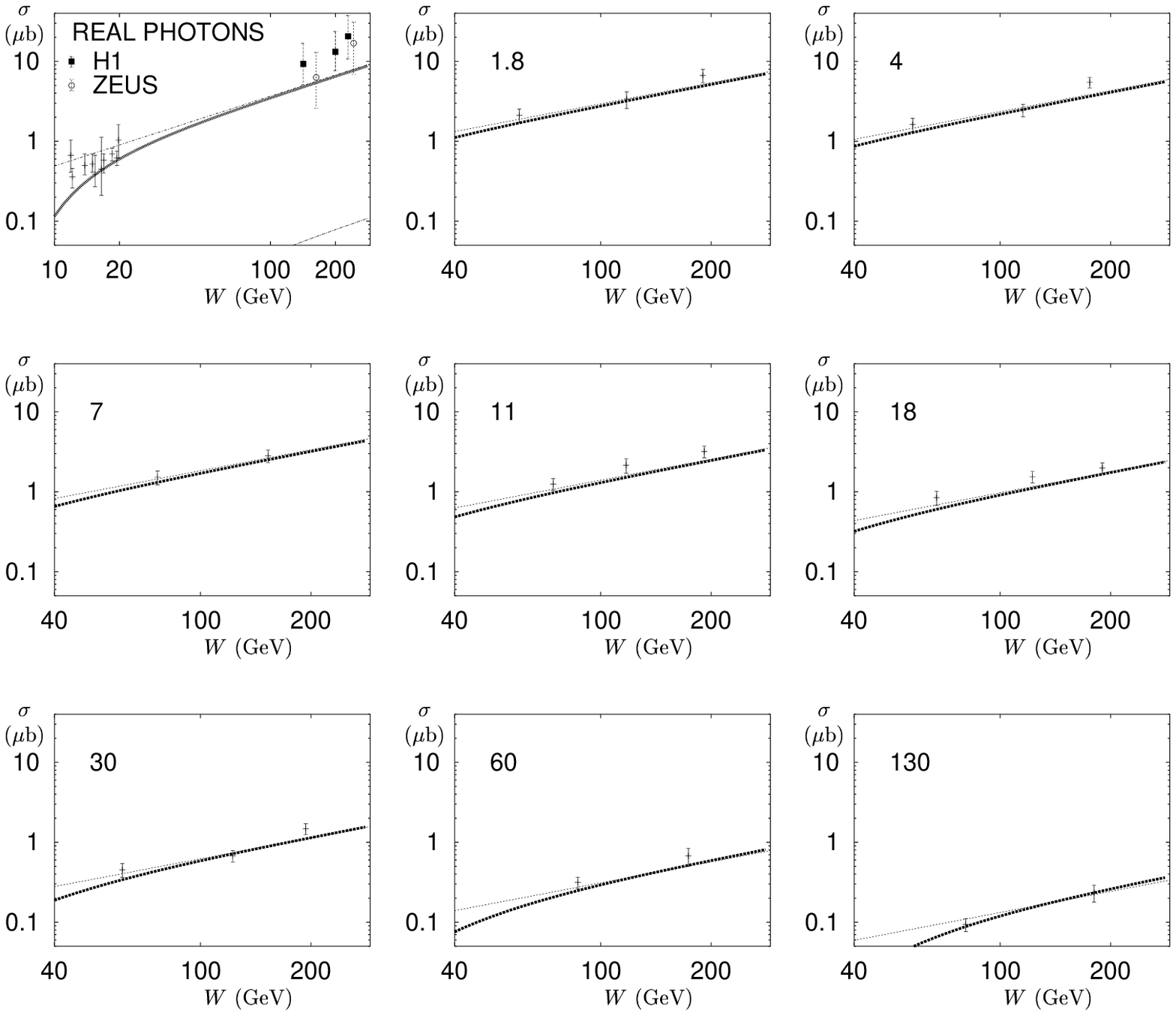}
\vskip 1truemm
Figure 2: Charm cross section: pQCD calculation (thick lines)
and the form (2) (thin lines). The data for $Q^2>0$ are from
ZEUS\ref{\zeusc}.
The photoproduction data are from H1\defref\h1photo{
H1 collaboration: S Aid et al, Nuclear Physics B472 (1996) 32
}
and ZEUS\defref\zeusphoto{
ZEUS collaboration: M Derrick et al, Physics Letters B349 (1995) 225
},
who give references to the fixed-target data. The line in the lower
right-hand corner of the photoproduction plot is our calculation
for $b$-quark production (for mass 4 GeV).
\endinsert

The hard-pomeron contribution to the complete $F_2(x,Q^2)$ is the same,
with $A_c$ replaced with\break\hbox{$A\approx 1.5\times 10^{-3}$.}
We have shown\defref\evol{
A Donnachie and P V Landshoff, Physics Letters B533 (2002) 277
}
that DGLAP evolution, with a gluon structure function that is dominated
at small $x$ by hard-pomeron exchange alone, 
produces a $Q^2$ dependence for the hard-pomeron part of $F_2$ that
agrees numerically with this.
Our procedure, which can be applied only to the hard-pomeron component,
gave almost identical outputs for LO and NLO evolution.
A good numerical fit to
the output of the DGLAP evolution for the small-$x$ behaviour of
the gluon structure function is
$$
xg(x,Q^2) \sim 0.95\,(Q^2)^{1+\epsilon_0}(1+Q^2/0.5)^{-1-\epsilon_0/2}\, x^{-\epsilon_0}
\eqno(3)
$$
This fit is valid for $Q^2$ between 5 and 500 GeV$^2$.

We use the gluon distribution (3) to calculate charm production in
leading-order pQCD and compare the result with (1). First, we calculate simple
photon-gluon fusion in lowest order, for which the relevant equation
\defref\roberts{
R G Roberts, {\sl The Structure of the Proton}, Cambridge University
Press (1990)
} is (5.112) of the book by Roberts\footnote{$^*$}{Note that the formula
Roberts gives for $v$ should be for $v^2$.}
$$
F_2^c(x,Q^2)=2 e_c^2{\alpha_s(Q^2+4m_c^2)\over2\pi}\int _{ax}^1dy\,g(y,Q^2+4m_c^2)f(x/y,Q^2)
\eqno(4a)
$$
where
\par\penalty-500
$$
f(z,Q^2)=v\Big[4z^2(1-z)-\half z -{2m_c^2\over Q^2}z^2(1-z)\Big]
+\Big[\half z -z^2(1-z)+{2m_c^2\over Q^2}z^2(1-3z)-{4m_c^4\over Q^4}z^3\Big]L
$$$$
a=1+4m_c^2/Q^2
$$$$
v^2=1-4m_c^2/[Q^2(y/x-1)]
$$$$
L=\log\Big({1+v\over 1-v}\Big)
\eqno(4b)
$$
For this leading-order calculation we again\ref{\evol} set
$\Lambda_{\hbox{{\sevenrm QCD}}}=140$ MeV.
We have to choose the argument of $\alpha_s$ and the scale $\mu$ of the
gluon structure function; physical intuition leads us to take
$Q^2+4m_c^2$ for both, though it must be recognised that this is a mere
guess. We need also to fix a value for $m_c$. We find that 1.3 GeV
gives good results: the thick lines in figure 2 are the output of the
calculation, while the thin lines are the phenomenological fit (2).

As we have said, our gluon distribution is hard-pomeron dominated; when
we use it to calculate charm production we are modelling the strength of
the coupling of the hard pomeron to the charm quark.
As can be seen from the plots in figure 2, the calculation also includes
threshold effects which make the rise steeper than $W^{2\epsilon_0}$ at
small $W$. To a small extent, these threshold effects depend on
the behaviour of the gluon distribution for values of $x$ that are
beyond the small-$x$ region where (3) is valid. We have used (3)
multiplied by $(1-x)^n$ with $n=5$; changing $n$ by one unit changes the
charm photoproduction cross section by less than 15\% at $W=10$~GeV
and by 1\% or less when $W> 50$~GeV. 

At low $Q^2$, and particularly for photoproduction, the magnitude
of the cross section is sensitive to the value chosen for $m_c$.
Changing $m_c$ by 100 MeV away from our preferred value of 1.3 GeV
changes the photoproduction cross section by more than 20\% at the
higher energies.
We have not included any possible contribution from the hadronic
structure of the photon, because its magnitude is so uncertain.
Our calculations are consistent with it being small, but this may
not be true\defref\frix{
S Frixione et al, Adv Ser Direct High Energy Phys  15 (1998) 609
(hep-ph/9702287)
}.
If indeed it is small, one needs\ref{\frix} a structure function such as ours
or the old MRSG\defref\mrsg{
A D Martin, R G Roberts and W J Stirling, Physics Letters B354 (1995) 155
}
to reproduce the steep $W$-dependence of the data\ref{\frix} at
small $Q^2$. The more modern
MRST2001\ref{\mrs}
and CTEQ6M\ref{\cteq}
gluon structure functions are not large enough, and not
steep enough, to reproduce the low-$Q^2$ data, as is obvious from figures 
1 and 3. 

\topinsert
\centerline{\epsfxsize=0.5\hsize\epsfbox[50 50 400 300]{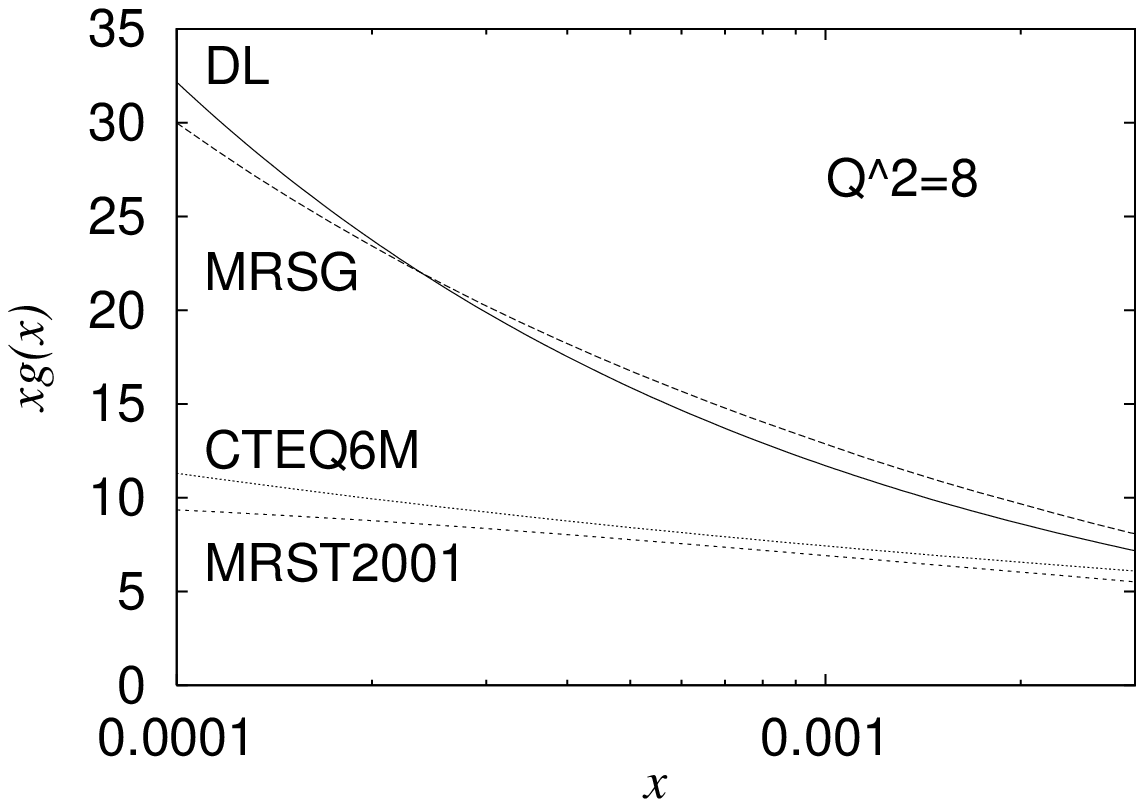}}
\vskip 1truemm
Figure 3: Various gluon structure functions\defref\durham{
Durham data base, http://cpt19.dur.ac.uk/hepdata/pdf3.html
}
at $Q^2=8$ GeV$^2$
\vskip 5truemm
\endinsert

So, with $\mu^2=Q^2+4m_c^2$ and $m_c=1.3$ GeV, leading-order
photn-gluon fusion gives a good description of the data for $F_2^c$
with hard-pomeron exchange alone, even down to $Q^=0$. At small enough $x$,
the corresponding $c$-quark density is found to be almost identical 
with the hard-pomeron components of the densities of the light quarks.
We previously\ref{\twopom} extracted the latter from the data for
$F_2(x,Q^2)$ and\ref{\evol} showed that, for $Q^2$ greater than about
5 GeV$^2$, they agree very well with 4-flavour zero-mass DGLAP evolution.
It is standard\defref\acot{
M A G Aivazis, J C Collins, F I Olness and W-K Tung, Physical Review
D50 (1994) 3102
}\defref\robthorne{
R S Thorne and R G Roberts, Physics Letters B421 (1998) 303
}\defref\smith{
M Buza, Y Matiounine, J Smith and  W L van Neerven, European Physical
Journal C1 (1998) 301
} 
that photon-gluon fusion at small $Q^2$ must be matched to DGLAP evolution
at large $Q^2$. Our calculation has this property: the two agree over a
large range of $Q^2$ values.

We have not attempted to make a best fit to the charm-production data.
In this, our approach is different from the so-called global 
fits\ref{\mrs,\cteq}. 
Rather, our emphasis is on simplicity; we used only three free parameters
in the fit to the small-$x$ behaviour of $F_2(x,Q^2)$, and 
introduced no additional ones for $F_2^c(x,Q^2)$. The physics underlying
our approach is very different from the conventional one and our 
successful application of pQCD to $F_2^c(x,Q^2)$ is a striking
confirmation of the correctness of our extraction of the hard component
from $F_2(x,Q^2)$. Most of $F_2(x,Q^2)$ at small $x$ is the contribution
from the soft component; this therefore plays the dominant role in
the conventional approach, but it has not entered at all into the analysis
we have described here.

The first plot in figure 2 shows also our calculation for the $b$-quark
photoproduction cross section using $m_b = 4$ GeV. It is not inconsistent
with a measurement of H1\defref\h1b{
H1 collaboration: C Adloff et al, Physics Letters B467 (1999) 156}.

We now consider the proton's longitudinal structure function
$F_L(x,Q^2)$. We calculate this in leading-order pQCD. The relevant
equation is (5.110) of the book by Roberts\ref{\roberts}. However, we
include the effect of the mass $m_c$ in the contribution from $c,\bar c$;
this correction may be found in equation (E.3) of the review
by Budnev et al\defref\budnev{V M Budnev et al, Physics Reports C15 (1974) 
181}. So
$$
F_L(x,Q^2)=G(x,Q^2)+{4\alpha_s(Q^2)\over 3\pi}\int _x^1{dy\over y}
\Big ({x\over y}\Big)^2F_2(y,Q^2)
\eqno(5)
$$
where the contribution of the charm quark to $G(x,Q^2)$ is
$$
G^c(x,Q^2) = 2e_c^2{\alpha_s(Q^2+4m_c^2)\over \pi}\int _{xa}^1dy\,
\Big ({x\over y}\Big)^2 \Big[\Big(1-{x\over y}\Big)v
-{2m_c^2x\over Q^2y}L\Big)\Big]\,g(y,Q^2+4m_c^2)
\eqno(6)
$$
with $a,v$ and $L$ defined in (4b).
We have again had to choose the argument of $\alpha_s$ and the
$Q^2$-scale of the
gluon structure function. We have made the same choice as before:
$Q^2+4m_c^2$ for each.
Again this is a guess and the output is sensitive
to it at low $Q^2$.  The light quarks
contribute similarly, with $m_c$ replaced with 0.

\pageinsert
\line{\hfill
\epsfysize=0.47\vsize\epsfbox[50 47 545 770]{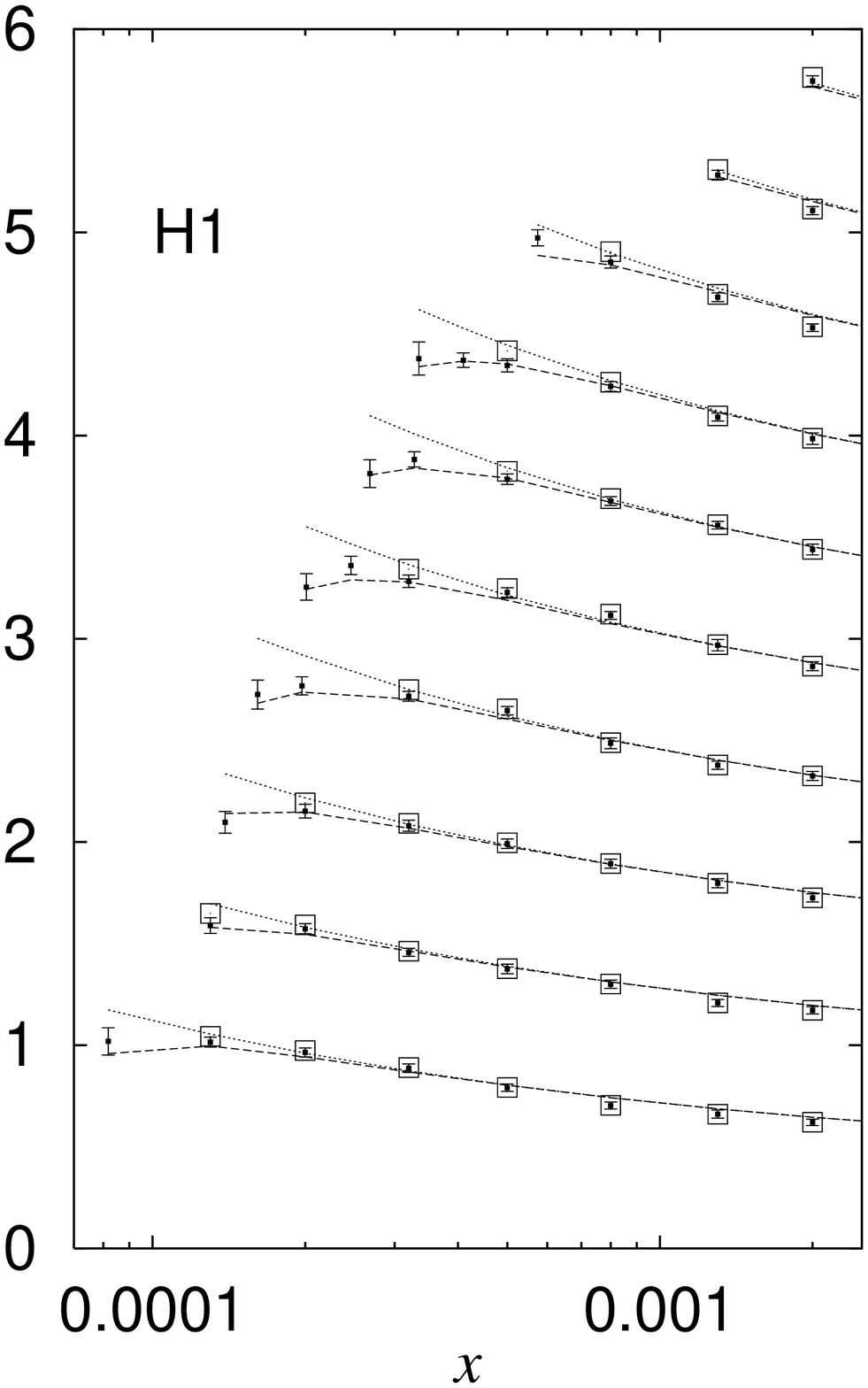}\hfill
\epsfysize=0.47\vsize\epsfbox[50 47 545 770]{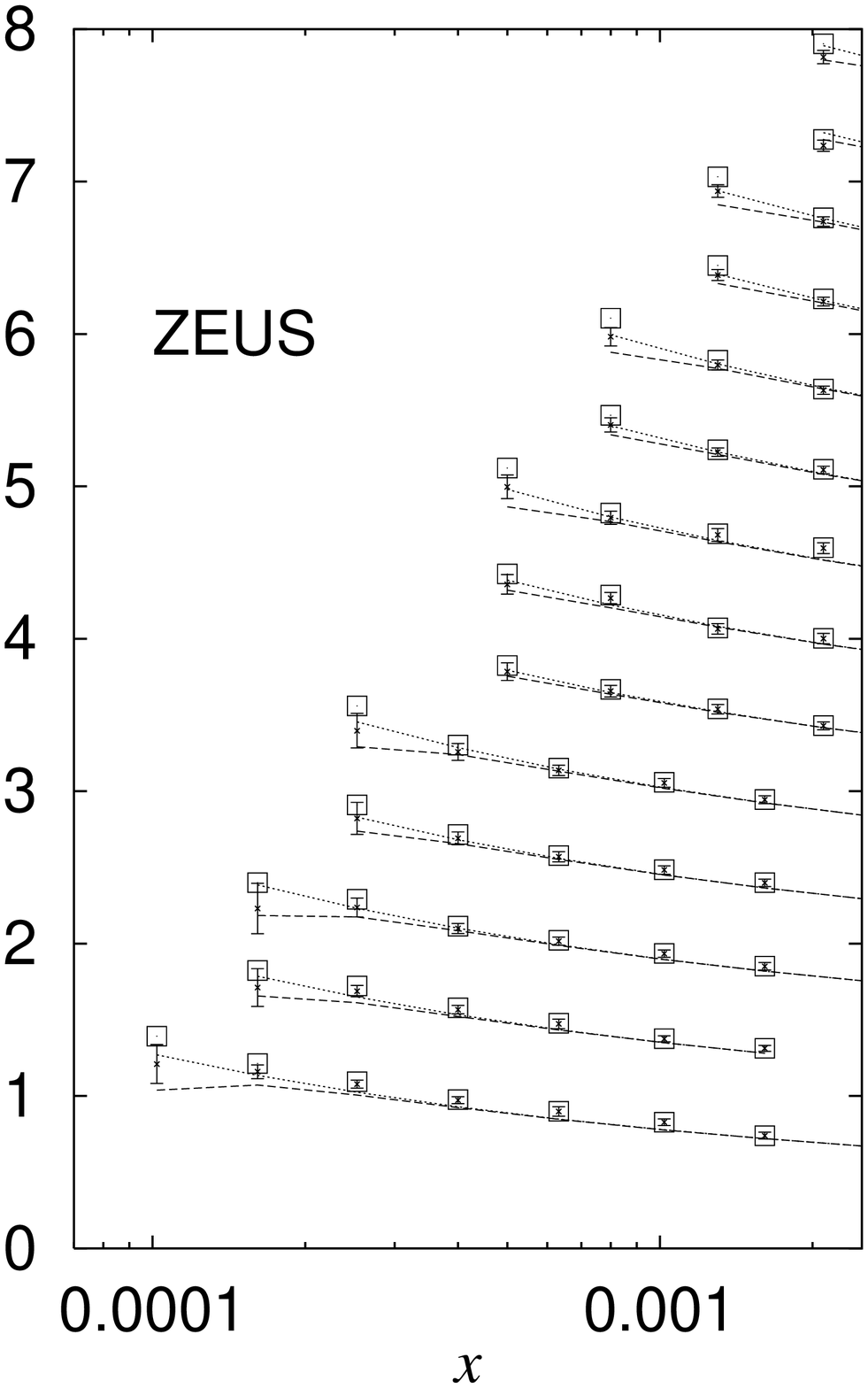}\hfill}
\vskip 0truemm
\line{\hfill (a) \hfill\hfill (b)\hfill}
\vskip 1truemm
Figure 4: H1\defref\hone{
H1 collaboration: C Adloff et al, European Physical Journal C21 (2001) 33
}
and ZEUS\defref\chek{
ZEUS collaboration: S Chekanov et al, European Physical Journal C21 (2001)
443
}
data for the reduced cross section (7). The open
squares are
the values of $F_2(x,Q^2)$ extracted from these data by the two experiments.
The data in (a) range from $Q^2=5$ GeV$^2$ at the bottom to 60 at the top
and in (b) from
6.5 to 120; in each case the data
are separated by adding an extra 0.5
at successive values of $Q^2$. At each $Q^2$ the upper line is our
fit\ref{\twopom} to $F_2(x,Q^2)$ (the ZEUS data in (b) were not available
when the fit was made) and the lower line is the calculated reduced
cross section.
\vskip 6truemm
\epsfysize=0.28\vsize\epsfbox[45 540 555 773]{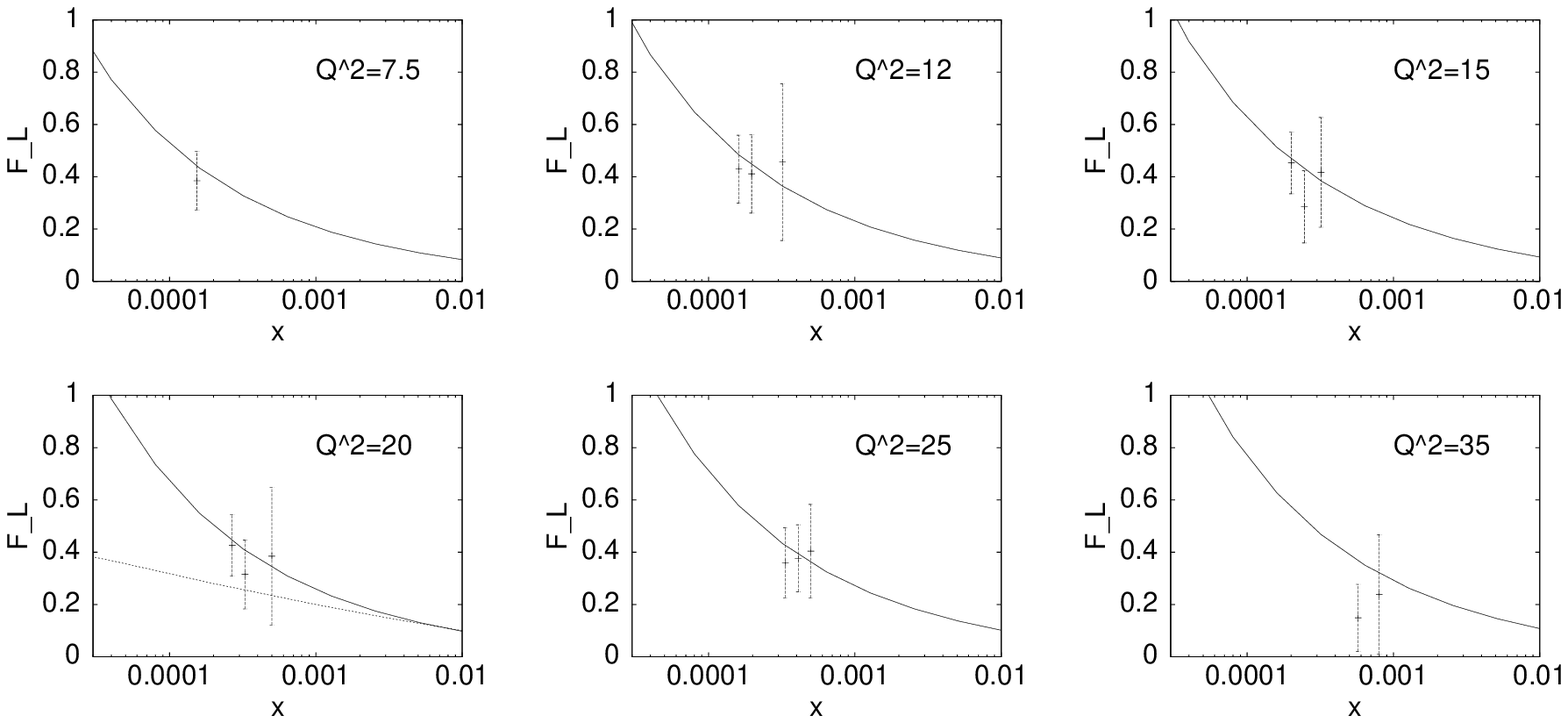}
\vskip 2truemm
Figure 5: H1 data\ref{\hone} for the longitudinal structure function
at various $Q^2$ values, with our pQCD calculations. The lower line
on the $Q^2=20$ GeV$^2$ plot is the MRST2001 prediction\ref{\mrs}.
\endinsert

The HERA experiments measure the reduced cross section
$$
\sigma^r(x,y,Q^2)=F_2(x,Q^2)-{y\over 1+(1-y)^2}F_L(x,Q^2)
\eqno(7)
$$
We have calculated this from our fit to $F_2$ and our gluon structure function
(3); the results are shown in figure 4. The ranges of $x$ and $Q^2$ shown
are chosen because our fit to $F_2$ used only small-$x$ data and because
we found\ref{\evol} that perturbative evolution only described it well
for $Q^2$ greater than about 5~GeV$^2$, so our gluon distribution is
not reliable for smaller values. In figure 5 we compare our
calculated $F_L$ with the values extracted by H1\ref{\hone}
from their data.  Separation of $F_L$
from $F_2$ requires extrapolation and depends on some assumed
parametrisation.

In conclusion, in this paper we have continued our programme of reconciling
the Regge and
pQCD-evolution approaches to structure function data. We use two sets of
data, the charm structure function and the longitudinal structure function.
In our calculation of $F_2^c$ photon-gluon fusion at small $Q^2$ is
matched to DGLAP evolution at large $Q^2$.
Our success in describing the charm and longitudinal structure functions 
provides further confirmation for the
correctness of the
two approaches and our understanding of how they fit together. We should
warn, however, that in each
case the extraction of the data from the raw measurements requires very large
extrapolations. As MRST have observed\ref{\mrs}, it would be particularly
useful to have good data for $F_L(x,Q^2)$, since this offers rather
direct information about the gluon distribution.

Our gluon structure function is larger and steeper
at small values of $x$ than is conventionally believed\ref{\mrs}\ref{\cteq},
particularly at small $Q^2$, but we have shown in this paper that
there is experimental support for it. A less-steep gluon distribution will
not explain the charm data at small $Q^2$.
Because our evolution procedure does not introduce spurious singulariies
at $N=0$ into the splitting matrix, we conclude that the gluon distribution
is larger than is usual in order to achieve the observed evolution of 
$F_2(x,Q^2)$.

\bigskip{\eightit
This research is supported in part by the EU Programme
``Training and Mobility of Researchers", Network
``Quantum Chromodynamics and the Deep Structure of
Elementary Particles'' (contract FMRX-CT98-0194),
and by PPARC}
\vskip 15truemm
\vfill\eject
\medskip\immediate\closeout\rfile\writestoppt
\baselineskip=10pt{{\bf References}}\bigskip{\frenchspacing%
\parindent=20pt\escapechar=` \input refs.tmp\bigskip}\nonfrenchspacing
\bye